\documentstyle[psfig,referee]{mn}

\title[GLITP optical monitoring of QSO 0957+561]{GLITP optical monitoring of 
QSO 0957+561:\\ 
$VR$ light curves and variability}
\author[Ull\'an et al.]
      {A. Ull\'an$^1$, L.\ J. Goicoechea$^1$, J.\ A. Mu\~noz$^{2}$, E. 
        Mediavilla$^3$, 
	\newauthor
	M. Serra-Ricart$^3$, E. Puga$^{3,4}$, D. Alcalde$^3$, A. Oscoz$^3$ and 
	R. Barrena$^3$ \\
	$^1$ Departamento de F\'{\i}sica Moderna, Universidad de Cantabria,
Avda. Los Castros s/n, E-39005 Santander, Spain \\ E-mail: 
aurora.ullan@postgrado.unican.es, goicol@unican.es \\
      $^2$ Departament d'Astronomia i Astrof\'{\i}sica, Universidad de Valencia,
Dr. Moliner 50, E-46100 Burjassot, Spain \\ E-mail: jmunoz@uv.es \\
	$^3$ Instituto de Astrof\'{\i}sica de Canarias,  Via L\'{a}ctea s/n, 
E-38200 La Laguna, Spain \\ E-mail: emg@ll.iac.es, mserra@ot.iac.es, 
dalcalde@ll.iac.es, aoscoz@ll.iac.es, rbarrena@ll.iac.es \\
	$^4$ Max Planck Institut f\"{u}r Astronomie, K\"onigstuhl 17, Heidelberg, 
Germany \\ E-mail: puga@mpia-hd.mpg.de}

\begin{document}

\maketitle

\begin{abstract}
The GLITP collaboration observed the first gravitational lens system (QSO 
0957+561) from 2000 February 3 to 2000 March 31. The daily $VR$ observations 
were made with the 2.56-m Nordic Optical Telescope at Roque de los Muchachos 
Observatory, La Palma (Spain). We have derived detailed and robust $VR$ light
curves of the two components Q0957+561A and Q0957+561B. In spite of the 
excellent sampling rate, we have not found evidence in favor of true daily 
variability. With respect to variability on time-scales of several weeks, we 
measure $VR$ gradients of about $-0.8$ mmag/day in Q0957+561A and $+ 0.3$ 
mmag/day in Q0957+561B. The gradients are very probably originated in the far 
source, thus adopting this reasonable hypothesis (intrinsic variability), 
we compare them to the expected gradients during the evolution of a compact 
supernova remnant at the redshift of the source quasar. The starburst scenario
is roughly consistent with some former events, but the new gradients do not
seem to be caused by supernova remnant activity. 
    
\end{abstract}

\begin{keywords}
galaxies: photometry -- gravitational lensing -- quasars: general -- 
quasars: individual: QSO 0957+561
\end{keywords}

\section{Introduction}

In some aspects the first lensed quasar QSO 0957+561 is a relatively enigmatic 
system. Although there is an agreement on the range for the optical time 
delay between its components $\Delta t_{BA}$, and currently, a rough interval of
415--430 days seems incontrovertible (Pelt et al. 1996; Kundi\'c et al. 1997; 
Oscoz et al. 1997; Pijpers 1997; Pelt et al. 1998; Serra-Ricart et al. 1999; 
Oscoz et al. 2001; Slavcheva-Mihova, Oknyanskij \& Mihov 2001; Ovaldsen et al. 
2003a), we have clear evidence for different delays associated with different pairs 
of twin intrinsic events (Goicoechea 2002). The light curves of both images in this 
gravitational mirage show variability on very different timescales, but when one 
concentrates on the well-sampled intrinsic events with an amplitude of about 100 
mmag and lasting several months, there are detected three different delays. The three 
delays between twin events are basically included in the previously quoted interval. 
In principle, the presence of multiple delays could indicate that local and violent 
physical phenomena (flares) are taking place in a source with finite size, and thus, 
the time delay distribution may be a basic tool to discuss the size and nature of 
the region of flares (Yonehara 1999). The existence of an extended region of flares 
implies that the optical source in QSO 0957+561 could be made of the standard engine
(accretion disc around a supermassive black hole) and at least other structure. A 
circumnuclear stellar region with starburst activity, a second accretion disc (binary
black hole) or jets with optical activity are good candidates to be a companion 
structure of the standard one. These findings encourage to carry out new monitoring 
campaigns of several months per year, which must be useful to find new pairs of twin 
intrinsic events and to map the positions of the flares in the quasar. Even a multiband 
monitoring during only a few months may be important, since the hypothetical presence 
of a prominent event (very probably caused by an intrinsic flare; see here below) 
would serve to analyze the duration, the released energy, and the origin of the flare 
associated with it. For example, Collier (2001) suggested that comparing the event in 
two optical bands, one could obtain a non-zero chromatic lag of $\sim$ 1 day 
(supporting the reverberation within an accretion disc), or maybe, an accurate zero 
lag (indicating the absence of disc reprocessing). To be successful with these studies, 
very precise and well-sampled brightness records are needful. 

The existence of rapid and very rapid microlensing variability has also been  
studied for several authors. At present the IAC group has carefully analyzed four 
difference light curves corresponding to the 1996, 1997, 1998, 1999, and 2000 
observing seasons in the $R$ band. The observations were made with the 0.82-m IAC-80
telescope at the Teide Observatory, and the difference light curves showed noisy 
behaviours around the zero line and no rapid (with a duration of months) events (see 
Gil-Merino et al. 2001 for details on the first two difference curves). In fact all 
the IAC difference signal can be due to observational noise. These conclusions 
agree with the results derived from the $g$-band photometric measurements at the 
Apache Point Observatory (APO) 3.5-m telescope for the period 1995-1996 (Kundi\'c et 
al. 1995; 1997). Schmidt \& Wambsganss (1998) did not find reliable microlensing 
imprints in the light curves by Kundi\'c et al. (1995, 1997). Therefore, there is a 
strong evidence against the existence of rapid microlensing in the components of the 
system, and very probably, all the features on a few months timescale are originated in 
the source quasar. Only the studies based on the CfA frames (CCD images taken with the 
1.2-m telescope at Fred Lawrence Whipple Observatory) disagree with this point of view. 
However, the rapid "microlensing" events found by Schild (1996) (see also Ovaldsen et 
al. 2003a) could be related to either some kind of observational noise (underestimation 
of errors) or the assumption of a unique delay. We note that the 1.2-m telescope on Mt. 
Hopkins, Arizona, did not work in very good conditions: angular resolution of about 
0.65 arcsec/pixel, mean seeing value (FWHM) of around 2 arcsec, and PSFs with coma-like 
appearance (Ovaldsen et al. 2003a). On the other hand, the assumption of a unique delay 
does not seem suitable in the analysis of large records including several features. 
In fact, using small segments of the whole records, some probes suggested the existence 
of three well-separated delays (Ovaldsen, private communication). The very rapid 
microlensing events (with timescales ranging from a few days to a few weeks) and the 
ultrarapid extrinsic events, which were reported by Schild and collaborators (e.g., 
Schild 1996, Colley \& Schild 2003), are more subtle than the rapid ones and 
as far as we know they have not been rejected/confirmed yet. In order to discuss
this topic in a proper way, both the very accurate knowledge of the involved time 
delays and excellent photometric data are required. 

The Gravitational Lenses International Time Project (GLITP) is a program to observe,
analyze, and interpret gravitational mirages and related objects. In particular,
the optical monitoring (OM) subproject focus on the light curves of the systems QSO 
0957+561 and QSO 2237+0305. In this paper (Sections 2 and 3) we present new $VR$ 
light curves for the two components A--B of QSO 0957+561 ($z_s$ = 1.41). The 
GLITP-OM/Q0957+561 project was conceived (amongst other things) to search for gradients 
in the light curves of the system, so the study by Ovaldsen et al. (2003b) and our 
effort are complementary works. Ovaldsen et al. (2003b) discussed the hourly and daily 
variability, while we address the daily, weekly and monthly variability (Section 3). 
In Section 4, we compare the observed $V$-band gradients and the expected ones after a 
supernova explosion inside a high-density medium at $z$ = $z_s$ = 1.41. Finally, Section 
5 summarizes our main conclusions.

\section{$VR$ observations and first light curves}

We observed QSO 0957+561 from 2000 February 3 to 2000 March 31, i.e., during two 
months in 2000. All observations were made with the 2.56-m Nordic Optical Telescope 
(NOT) at the Roque de los Muchachos Observatory, Canary Islands (Spain). The images 
were taken with StanCam, a camera which uses a SITe 1024$\times$1024 CCD detector 
with a 0.176 arcsec/pixel scale. In the observing season, exposures in the $V$ and $R$ 
filters were taken every other night when clear. For each monitoring night, in general, 
we have three consecutive exposures, i.e., one 300 s exposure in the $V$ passband and 
two 150 s exposures in the $R$ passband. As usual the preprocessing of the data included 
bias subtraction and flat-fielding using sky flats.

\begin{figure}
\psfig{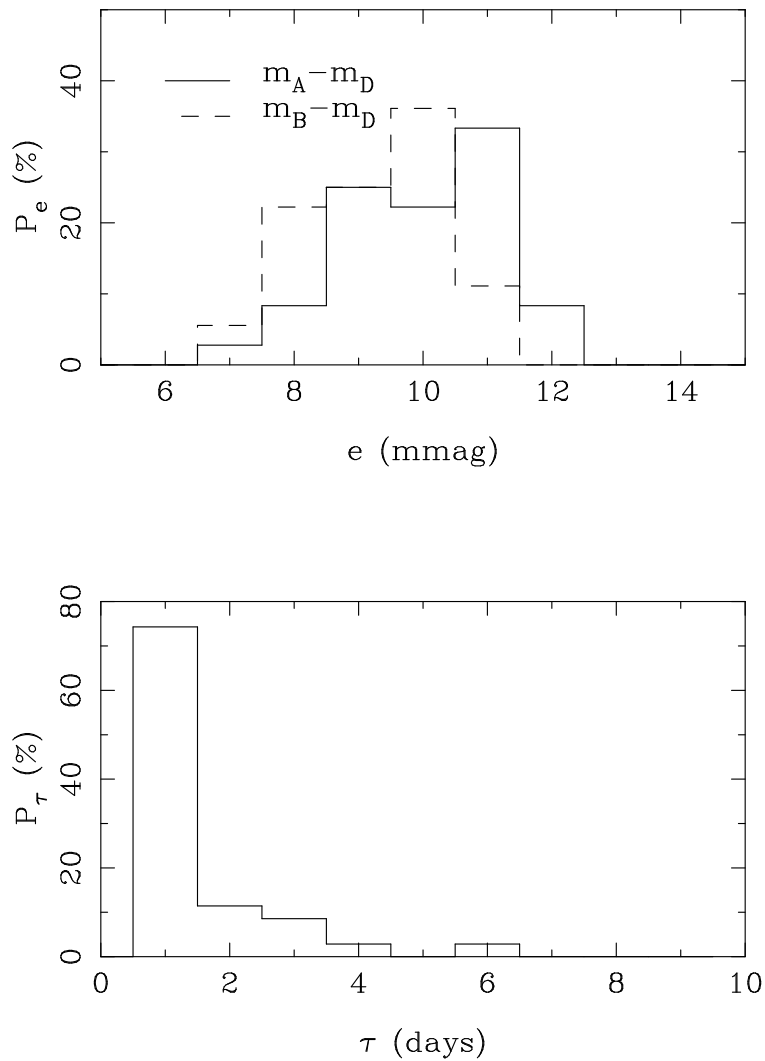}  
\caption{Some properties of the $R$-band data from {\it pho2com}: probability 
distribution of formal errors in $m_A - m_D$ and $m_B - m_D$ (top panel), and 
probability distribution of time separations between adjacent data (bottom panel).
We remark that the mean error is of $\approx$ 10 mmag and the adjacent data are 
mostly separated by only one day.}
\label{Fig. 1}
\end{figure}

\begin{figure}
\psfig{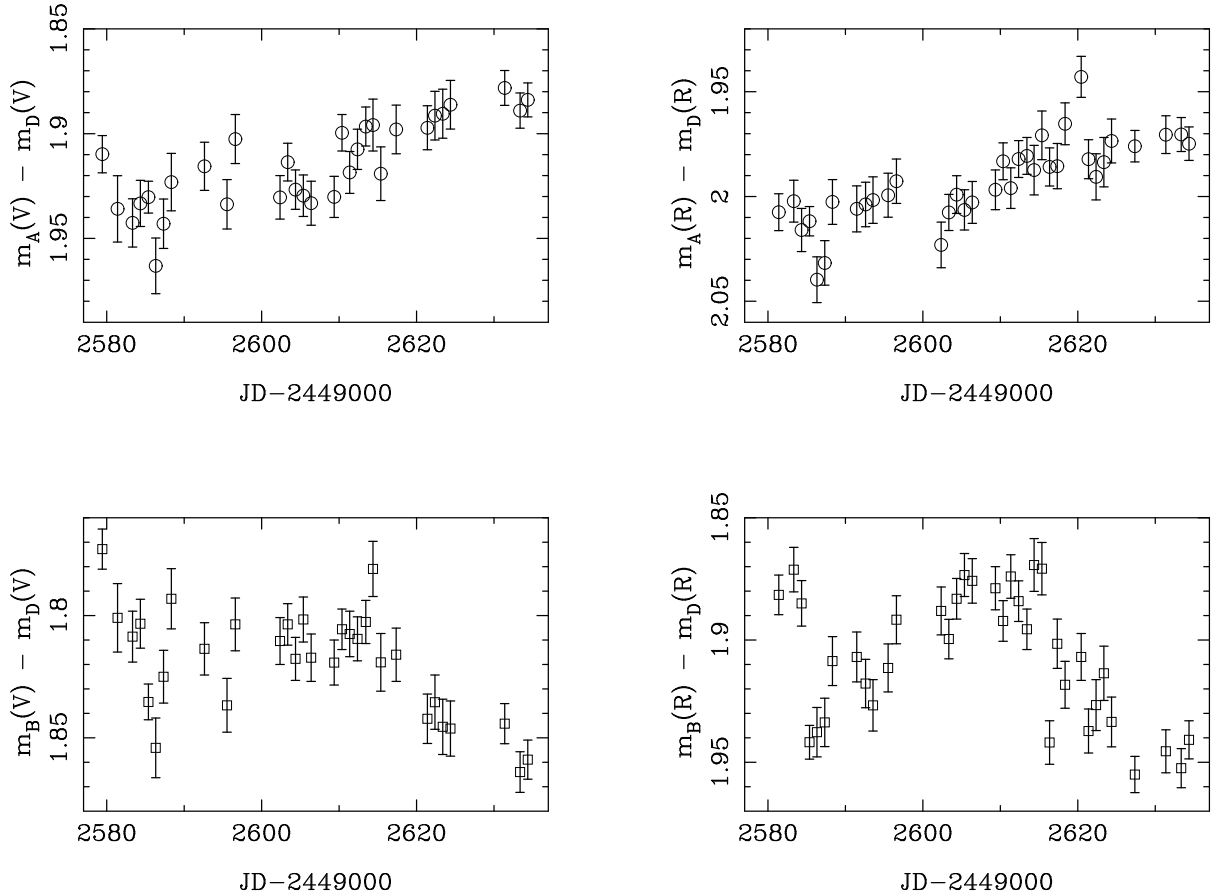}
\caption{$VR$ light curves from {\it pho2com}. The open circles trace the light
curves of Q0957+561A, while the open squares draw the brightness records of
Q0957+561B. We see a rise in the flux of Q0957+561A and an unclear behaviour of
Q0957+561B. In the right-hand bottom panel, it appears an "event" from day 2585 to 
day 2630.}
\label{Fig. 2}
\end{figure}

To obtain the $VR$ light curves of Q0957+561A and Q0957+561B, we use two different 
data processing techniques. In this section, we focus on a first photometry from the 
{\it pho2com} task. In a given optical band, from the {\it pho2com} technique we can 
infer the difference, in terms of magnitude, of the A--B components to a selected field 
star (for details on the whole procedure and the field stars, see Serra-Ricart et al. 
1999). The D star is chosen as the reference candle, and so, brightness records $m_A - 
m_D$ and $m_B - m_D$ are derived and properly analyzed. The mean FWHM of the seeing 
disc was below 1\arcsec\ for about 35--40\% of nights in the $V$ and $R$ bands. The 
initial $V$-band light curves show a rare behaviour around 2000 February 18 and 2000 
March 14, which is similar in both components. Therefore, to avoid the presence of 
artifacts, we filter the initial $V$ brightness records by dropping the fluxes that 
are in strong disagreement with the two adjacent data, and provided that the discrepancies 
simultaneously occur in both components. Our filtering criterion is simple: "strong
disagreement" means a difference exceeding three times the photometric error and
"adjacent data" means the previous and subsequent fluxes, when they are situated within
one week of the flux of interest. The procedure leads to 31 good data for each 
component in the $V$ filter (6 data are dropped in the filtering process). The $R$-band 
photometry does not show any anomaly, and in consequence, we consider the 36 initial 
red data for each component as the red dataset. To test the quality of the $R$-band 
monitoring, in Fig. 1 we depict the distribution of formal errors in $m_A - m_D$ and 
$m_B - m_D$ (top panel), and the distribution of time separations between adjacent data 
(bottom panel). We note that the mean error is of about 10 mmag and the sampling rate is 
excellent. In Figure 2, we show the light curves for QSO 0957+561: Q0957+561A in the $V$ 
and $R$ bands (left-hand and right-hand top panels, respectively) and Q0957+561B in the 
$V$ and $R$ bands (left-hand and right-hand bottom panels, respectively). As it was 
discussed in Fig. 5 of Serra-Ricart et al. (1999), the subtraction of the lens galaxy is 
not perfect, and some of its light could be still present in the $m_B - m_D$ fluxes. 
Therefore, the true $m_B - m_D$ fluxes may be out of the formal error bars, and it 
should be considered an additional correction. The contribution of the residual galactic 
light can be derived from simulations (Serra-Ricart et al. 1999) or from another data 
processing technique. Thus, to take into account the residual galactic light and check 
the reliability of the $m_A - m_D$ trends, we also use a PSF fitting code: the {\it 
psfphot} task. This PSF photometry scheme is applied in the next section.

As a reminder, we note that the {\it pho2com} task combines aperture photometry for 
reference stars and PSF fitting for reference stars and the two QSO components. First, 
the reference star fluxes are extracted through aperture photometry with a variable 
aperture of radius 2$\times$FWHM. Second, PSF fitting photometry, within a circle of 
radius FWHM, is applied to all the objects. Third, aperture corrections are computed 
from the previous data to compare the QSO component fluxes (aperture of radius 
2$\times$FWHM) with the reference star ones. This photometry code only attempts to fit 
the brightest region of the two QSO components, so that the lens galaxy is not taken into 
account. Therefore, if one QSO component is near to the lens galaxy and it is affected by 
galactic light, then the {\it pho2com} fluxes of the contaminated component will be 
overestimated. Moreover, both the galaxy/component confusion and the bias in the component 
flux will depend on the seeing (FWHM). This last fact was clearly proved in Serra-Ricart 
et al., and using the results from PSF fitting photometry, it is tested in the next 
section.

\section{Final light curves and variability}

In order to get final and robust $VR$ light curves of the two components Q0957+561A 
and Q0957+561B, we apply the {\it psfphot} photometric method (e.g., McLeod et al. 
1998), which is useful for extracting clean QSO fluxes (free of 
background, cross-contamination and contamination by the galaxy light). Assuming a de 
Vaucouleurs profile convolved with a point-spread function (PSF) as the observed lens 
galaxy profile, one can measure the brightness of the two quasar components through PSF 
fitting. The flux of the comparison star (D) can also be measured by means of PSF
fitting. Obviously, constant backgrounds are included to model the two regions of 
interest: the lens system (Q0957+561A + Q0957+561B + lens galaxy) and the D star. 
We use the clean two-dimensional profiles of three field stars (G, H and E) as 
empirical PSFs. Our framework is close to the methodology presented in Section 3 of 
Alcalde et al. (2002). Because of both the relatively faint bright of the elliptical 
lens galaxy in the frames and the proximity between the galaxy centre and the B 
component peak, we determine the relevant information on the galaxy from the best 
images (in terms of seeing values). Therefore, we apply the code to each image with a 
seeing (FWHM) better than 1.5 arcsec, using the brightest reference PSF and allowing 
all parameters to be free. For some frames, the method is not able to accurately 
extract several physical parameters of the lens galaxy, leading to results in apparent 
disagreement with the global distributions. Thus, in the estimation of each parameter, 
we follow a scheme with two steps. First, the values with a deviation (= value $-$ 
average) exceeding the standard one are dropped. Second, the parameter is inferred 
from the average of the "surviving" values. Finally, we obtain the morphological 
parameters of the galaxy (i.e., the effective radius, $R_{\rm eff}$, the ellipticity,
$\epsilon$, and the position angle, P.A.), the relative position of the galaxy 
(position relative to the A component) and the relative flux $\Gamma = f_{gal}/f_D$.
After, we apply the code to all images (whatever their seeings), setting the galaxy 
parameters to those derived in the previous step ($R_{\rm eff}$, $\epsilon$, P.A., 
relative position and $\Gamma$), using galaxy fluxes given by $f_{gal} = \Gamma \times 
f_D$ and allowing the remaining parameters to vary. In this last iteration, all the 
available PSFs are used. As the H star is relatively bright and it is present in most
frames, the brightness records from the PSF of the H star are regarded as the standard
ones.

\begin{figure}
\psfig{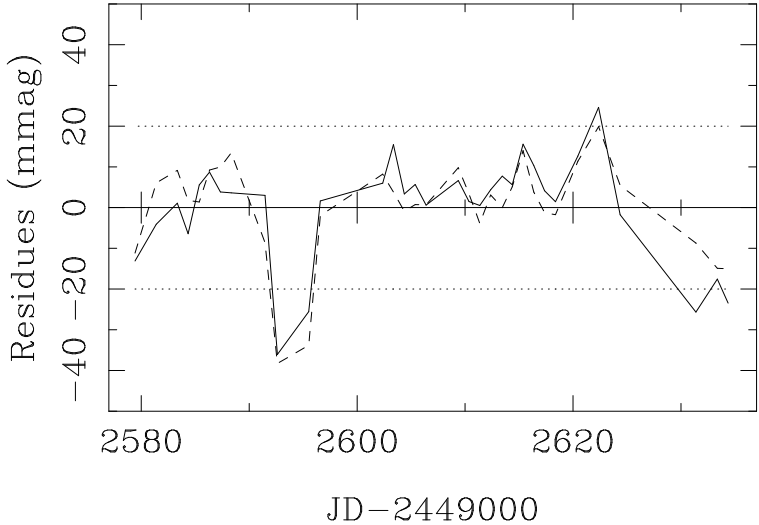}  
\caption{Standard residual curves in the $R$ band ({\it psfphot}). It is evident the 
strong correlation between the residues for Q0957+561A (solid line) and the residual 
signal for Q0957+561B (dashed line). A few prominent peaks exceed the $\pm$ 20 mmag 
levels (dotted lines).}
\label{Fig. 3}
\end{figure}

\begin{figure}
\psfig{figure=f4.eps,width=10cm}  
\caption{Standard $R$-band light curves from {\it psfphot} and the corresponding 
linear fits. The fluxes with very significant observational noise were not
taken into account.}
\label{Fig. 4}
\end{figure}

\begin{figure}
\psfig{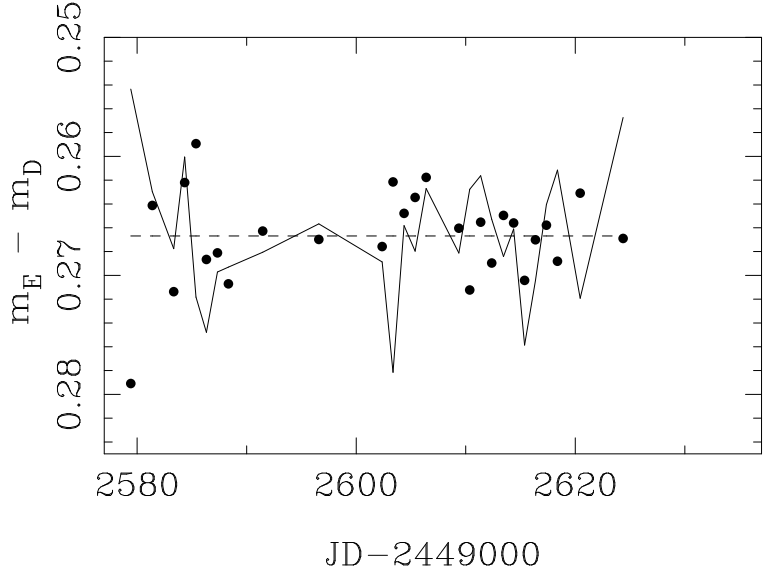}  
\caption{Standard $R$-band values of $m_E - m_D$ (filled circles) and standard $R$-band 
residues for Q0957+561A (solid line). We used the {\it psfphot} task, discarded the 
frames that lead to prominent peaks of observational noise and properly offset the
residual signal.}
\label{Fig. 5}
\end{figure}

\begin{figure}
\psfig{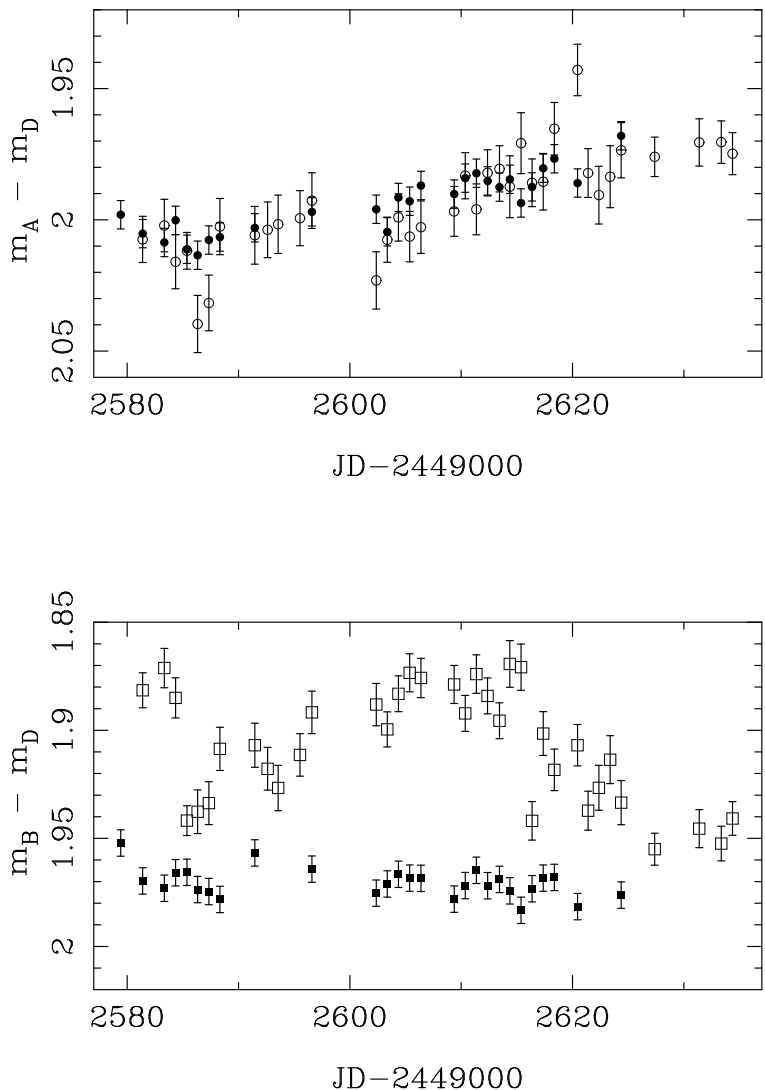}  
\caption{Comparison between the {\it pho2com} (open symbols) and {\it psfphot} (filled 
symbols) photometries in the $R$ band. We note the excellent agreement between the A 
fluxes from both techniques and the smooth behaviour of the B fluxes from {\it 
psfphot}. The residual galactic light is responsible for the "event" in the B 
component from {\it pho2com}.}
\label{Fig. 6}
\end{figure}

\begin{figure}
\psfig{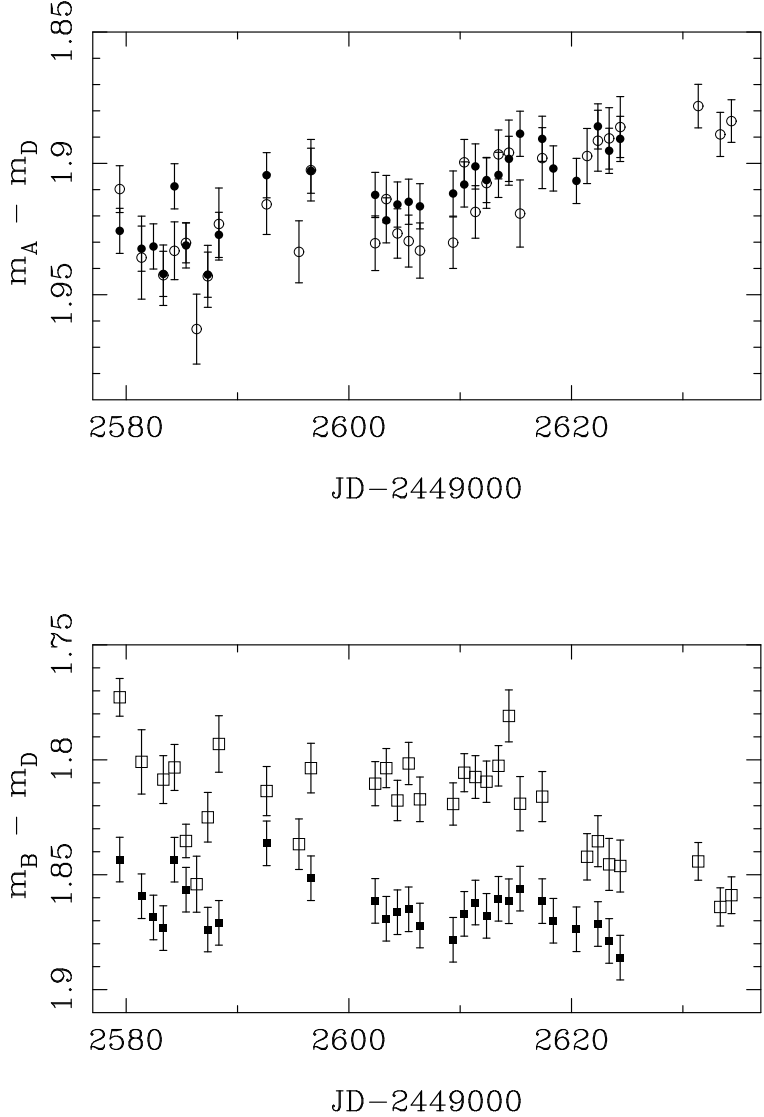}  
\caption{Comparison between the {\it pho2com} (open symbols) and {\it psfphot} (filled 
symbols) photometries in the $V$ band.}
\label{Fig. 7}
\end{figure}

\begin{figure}
\psfig{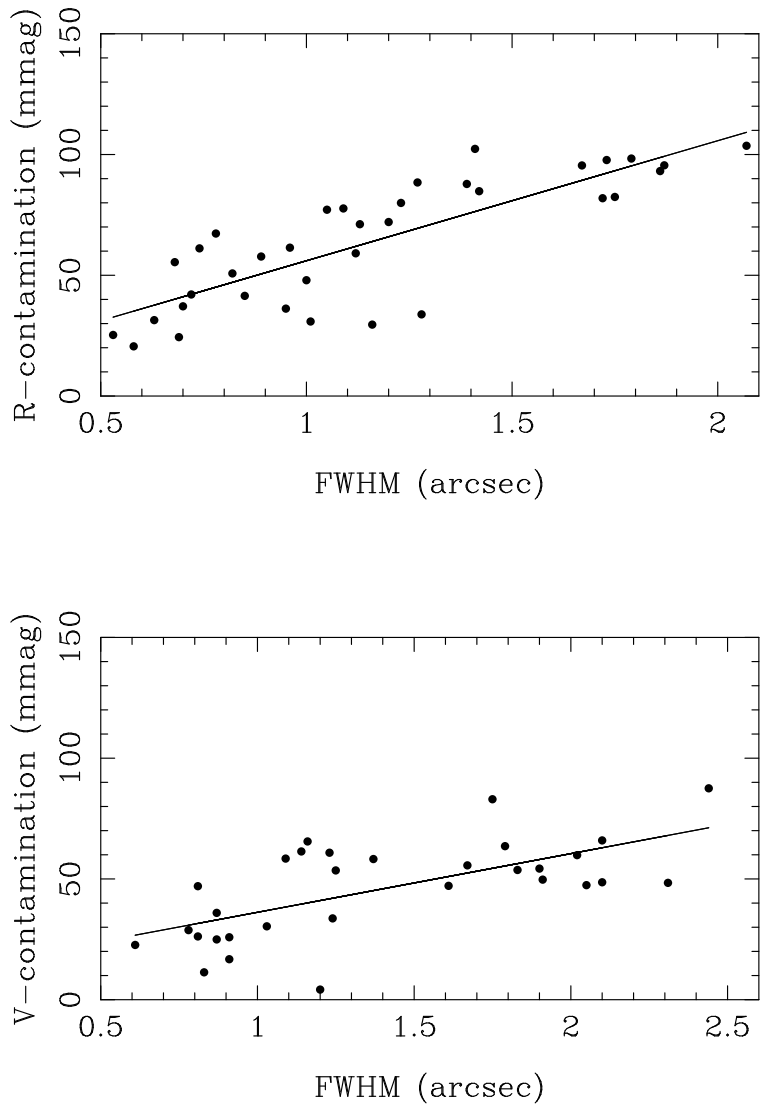}  
\caption{$R$-band and $V$-band contamination of the Q0957+561B/{\it pho2com} fluxes as 
function of seeing. In each optical band, the contaminations by galaxy light are estimated 
from the differences $C = y_B(psfphot) - y_B(pho2com)$. The linear fits $C = a\times$FWHM 
+ $b$ are also shown.}
\label{Fig. 8}
\end{figure}

\begin{figure}
\psfig{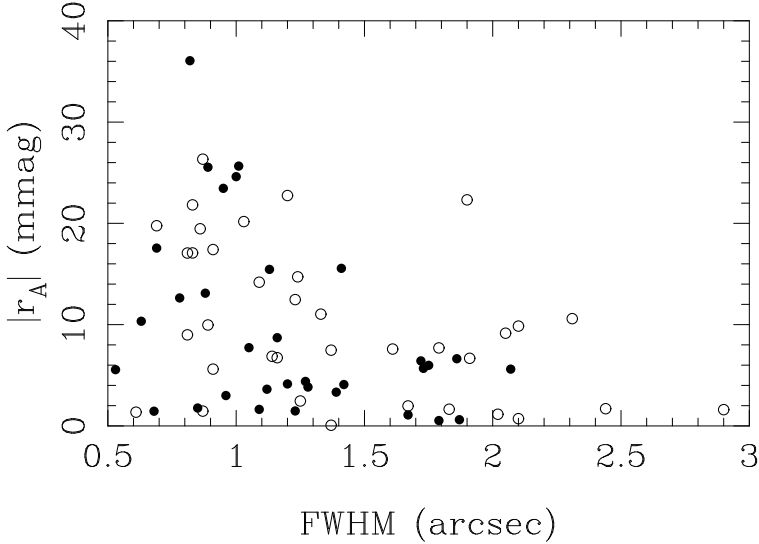}  
\caption{Amplitudes of the residues for Q0957+561A as a function of seeing (FWHM). The 
filled and open circles represent the $R$-band and $V$-band amplitudes, respectively. We 
used all the frames and the {\it psfphot} task.}
\label{Fig. 9}
\end{figure}
 
The standard $R$-band light curves $y_A = m_A - m_D$ and $y_B = m_B - m_D$ can be 
fitted to linear gradients $g_A(t)$ and $g_B(t)$, respectively, in such a way that
$y_A(t_j) = g_A(t_j) + r_A(t_j)$ and $y_B(t_j) = g_B(t_j) + r_B(t_j)$, where $t_j$
represent the observation dates, and $r_A$ and $r_B$ are residual signals. By doing the
fits, we derive two very different gradients $g_A$ and $g_B$, which reasonably describe
intrinsic variations in two different QSO epochs. Moreover, some to us surprise, we 
obtain very similar residues $r_A$ and $r_B$. These residual signals are drawn in 
Figure 3, where we see $r_A$ (solid line), $r_B$ (dashed line) and $\pm$ 20 mmag levels
(dotted lines). In the past, other authors have also found some zero-lag correlation 
between the variations of both components (e.g., Ovaldsen et al. 2003a, b and references 
therein). The $R$-band residues could be mainly originated by either observational 
systematic noise or physical phenomena inside the Milky Way. A physical phenomenon in the 
far source would be seen at a certain time in Q0957+561A and very much later in Q0957+561B 
(see Introduction). In our photometry, the observational noise seems to be the source of the 
residual variability. The prominent peaks in Fig. 3 correspond to good images in terms of 
seeing (FWHM $<$ 1.5 arcsec), but bad images in relation to the determination of the lens 
galaxy parameters (see here above). If the photometric method is not able to accurately 
extract the galaxy information, then it seems reasonable to find a peak of noise in the 
measurement of other quantities, e.g., the QSO fluxes. Hereafter, the $R$-band residual 
signals are assumed to be observational (systematic + random) noise. This hypothesis is 
also justified in the next paragraph. We wish to remark that the residues in Fig. 3 are not 
related to the galaxy model from superbGLITP images. In fact, in a first photometric stage, 
we used the relative astrometry and profile of the lens galaxy from HST (Hubble Space 
Telescope) observations. However, as we found a clear systematic in the residual signals, 
we thought that this systematic could be due to the external constraints, which were derived 
in other experiments. Hence, taking into account the high quality of the superbGLITP images, 
we finally decided to use a self-consistent photometry code that is only based on the GLITP 
experiment. Our final results are similar to the first ones obtained from HST constraints, 
so the systematic residues in Fig. 3 are not caused by the superbGLITP galaxy model. In the
$R$ band, we get $R_{\rm eff}$ = 3.8 $\pm$ 0.8 arcsec, $\epsilon$ = 0.32 $\pm$ 0.05 and 
P.A. = 58 $\pm$ 9 deg, while in the $V$ band, the galaxy parameters are $R_{\rm eff}$ = 3.0 
$\pm$ 1.0 arcsec, $\epsilon$ = 0.2 $\pm$ 0.2 and P.A. = 40 $\pm$ 21 deg. The HST galaxy 
ellipticity and position angle vary in the ranges 0.1--0.3 and 40--60 deg, respectively
(Bernstein et al. 1997).

By discarding the frames that lead to prominent peaks of noise in the two QSO components, 
we get a new standard photometry in the $R$ band. The new light curves are fitted to 
linear gradients and the residual signals are estimated again. In Figure 4 we see the 
new standard $R$-band light curves from {\it psfphot} together with the corresponding 
fits. For the A component, we measure a rise of $-$ 0.72 $\pm$ 0.08 mmag/day during six 
weeks, whereas for the B component, we derive a decrease of + 0.20 $\pm$ 0.09 mmag/day 
for the same period. The typical uncertainties are $e_A = \langle r_A^2 \rangle ^{1/2} 
\approx$ 5 mmag and $e_B = \langle r_B^2 \rangle ^{1/2} \approx$ 6 mmag. In order to 
test the reliability of our hypothesis (the residual signals represent the observational 
noise in the QSO fluxes), we also study the new standard $R$-band values of $y_E = m_E - 
m_D$. The E star is the faintest field star, although it is about 1.7 magnitudes brighter 
than the A component. The $y_E$ values (filled circles) and the $r_A$ residues (properly 
shifted in magnitude; solid line) are shown in Figure 5. Both trends have fluctuations of 
similar amplitude, and the rms of the $y_E - \langle y_E \rangle$ signal is $e_E \approx$ 
4 mmag. As the QSO components are fainter than the E star and they are placed on a crowded 
region of the images, the photometric errors $e_A$ and $e_B$ should be greater than $e_E$. 
So, we obtained a totally consistent result: $e_A$ and $e_B$ are 1--2 mmag above $e_E$. An
important fact to remark is that the new standard $R$-band records are stable against the 
change of the reference PSF. We also note the high accuracy of the PSF photometry in the 
$R$ band (QSO fluxes with uncertainties of $\sim$ 5 mmag), which permits to detect 
gradients less than 1 mmag/day. 

In Figure 6, it appears a comparison between the {\it pho2com} (open symbols) and {\it 
psfphot} (filled symbols) photometries in the $R$ band. At first sight, the behaviour 
of $y_A$({\it pho2com}) is very close to the evolution of $y_A$({\it psfphot}). From
a quantitative point of view, the mean relative deviation $\langle |y_A(pho2com) - 
y_A(psfphot)|/e_{dev} \rangle$ is of 0.9. On the other hand, there is no agreement between 
the $y_B$({\it pho2com}) and $y_B$({\it psfphot}) trends. The light curve $y_B$({\it 
pho2com}) is contaminated by residual galactic light, with a mean contamination of 
$\langle y_B(pho2com) - y_B(psfphot) \rangle \approx$ $-$ 70 mmag. We adopt the PSF 
fitting results as a final and robust photometry that includes clean fluxes of the B 
component as well as very reliable fluxes of the A component.

The standard $V$-band brightness records $y_A$ and $y_B$ can also be fitted to linear
gradients. From the fits, we infer results similar to the previous ones in the $R$
band, i.e., very different gradients and close residual signals. The rms of the 
residual signals is of 13--15 mmag, and we think that the $V$-band residues are due to
observational noise, just like the $R$-band ones. The proof of it is that two images
taken the same night (2000 February 4) lead to Q0957+561A fluxes separated by 22 mmag  
and Q0957+561B fluxes separated by 17 mmag, so the around 20 mmag intrahour deviations
(due to observational quasi-systematic noise) are comparable to the differences 
$y_A(t_j) - y_A(t_{j+1}) \approx r_A(t_j) - r_A(t_{j+1})$ and $y_B(t_j) - y_B(t_{j+1}) 
\approx r_B(t_j) - r_B(t_{j+1})$, where $t_j$ and $t_{j+1}$ are two consecutive 
observation nights. By discarding the frames with good seeing and poor imprints of the 
lens galaxy (the photometric method does not find the galaxy parameters in a right way; see 
here above), the Q0957+561A light curve is consistent with a rise of $-$ 0.89 $\pm$ 0.12 
mmag/day during six weeks, and the Q0957+561B record agrees with a decrease of + 0.36 
$\pm$ 0.13 mmag/day for a 45 days period. These $V$ gradients are marginally consistent 
with the $R$ ones (1$\sigma$ confidence levels), and we cannot assure the existence of 
chromatic variations. Moreover, the rms of the residual signals decrease up to $e_A 
\approx$ 9 mmag and $e_B \approx$ 10 mmag. We get again a very good photometry with 
uncertainties below 10 mmag level, although the $R$-band data are better than the 
$V$-band ones. The rms of $y_E - \langle y_E \rangle$ is of about 3 mmag, and consequently, 
$e_A$ and $e_B$ are three times the typical error in the E star flux. Along all this 
paragraph, we deal with a standard photometry (based on the PSF of the H star). However, 
a change in the reference PSF does not modify the photometric results (e.g., using the PSF 
of the E star instead of the standard PSF).

In Figure 7 we can see a comparison of the {\it psfphot} (filled circles and squares) 
and {\it pho2com} (open circles and squares) light curves in the $V$ band. Fig. 7 
reveals two important facts: (1) there are no significant deviations between the $y_A$ 
measurements from both methods, with $\langle |y_A(pho2com) - y_A(psfphot)|/e_{dev} 
\rangle$ = 0.7, and (2) the light curve $y_B$({\it psfphot}) seems to be free of 
contamination by the galaxy light, while $y_B$({\it pho2com}) has a mean contamination of 
$\langle y_B(pho2com) - y_B(psfphot) \rangle \approx$ $-$ 50 mmag. Thus, the {\it psfphot} 
dataset is our final (and robust) photometry in the $V$ band. The final $VR$ light curves 
are available on request.

Once we have $VR$ light curves of Q0957+561B from {\it pho2com} and another photometric 
procedure which extracts fluxes without contamination by galaxy light ({\it psfphot}),
it is possible to check the expected correlation between the contamination of the {\it 
pho2com} fluxes and the seeing (FWHM). In Fig. 8, the $R$-band and $V$-band contaminations 
are depicted as function of seeing. The expected linear trends (see Serra-Ricart et al.) 
are confirmed in this work. We obtain contamination laws (in magnitudes): $C_R = 
0.0497\times$FWHM + 0.0063 and $C_V = 0.0243\times$FWHM + 0.0119, that qualitatively agree 
with the Serra-Ricart et al.'s ones.   

Very recently, Ovaldsen et al. (2003b) presented $VR$ photometry of QSO 0957+561 from 
four consecutive nights of intensive monitoring at the NOT. They did not find clear 
fluctuations within each night, but reported day-to-day fluctuations of a few 
milli-magnitudes. The authors also discussed the zero-lag correlation between the A and 
B light curves, and did not obtain a fair conclusion about the nature of the observed 
variability: true (physical) variations or observational noise. Our photometry shows a 
daily variability which is very similar in the two QSO components, and we justified that 
the day-to-day variations are due to observational noise. However, in order to go further 
on, we can try to look for the origin of the noise, or in other words, to answer the 
question: which is the reason for the peaks of noise?. Using the first photometric results
(before to discard the {\it bad} frames), in Fig. 9, we examine the amplitudes of the 
residues for Q0957+561A as a function of seeing (FWHM). The filled and open circles 
represent the $R$-band and $V$-band amplitudes, respectively. In Fig. 9, as the seeing is 
worse, relatively small residues appear. Most the prominent peaks of noise correspond to 
nights with good seeing conditions. This probes that bad seeing conditions do not cause 
the peaks of noise. As it was noted at the end of Sec. 6 in Colley \& Schild (1999), the 
PSF may vary over the field because of the optics of the telescope and camera, and the 
variation may be more significant in good seeing conditions. Alternatively, the effective
PSF associated with each object could depend on the object flux, the background flux and
other factors, so the stellar PSFs would be different to the PSF of the components.
Therefore, we feel that a slight mismatch between the reference PSF and the PSF in the lens 
system region could be the responsible for the peaks of noise. This possibility can be 
tested from two images in the $R$ band, one leading to small noise (2000 March 12) and other 
corresponding to the maximum noise in Fig. 9 (2000 February 18). After subtracting the 
background in the pixels associated with the lens system, we analyze the lens system region. 
While in March 12, observed instrumental fluxes and modelled instrumental fluxes are very 
similar in both the central region and the tail of the QSO components, in February 18, 
there are clear differences between the observed fluxes and the modelled ones.

Apart from the day-to-day variability, uncorrelated gradients on several weeks time-scales 
are unambiguously detected, so we can reasonably consider these variations on longer 
time-scales as true fluctuations originated in the far quasar. In the next section, taking 
into account the observed gradients, we test the feasibility of a possible physical 
scenario.

\section{GLITP $V$-band gradients and supernova remnant activity}

In recent papers, using a starburst model, several authors tried to explain the optical 
variability of QSOs (Aretxaga, Cid Fernandes \& Terlevich 1997; Kawaguchi et al. 
1998; Hawkins 2002). The starburst model consists of a central stellar cluster whose 
luminosity comes from the stars, the type II supernova explosions and the activity 
of supernova remnants. The details on the $B$-band supernova(SN)/supernova remnant(SNR) 
luminosity curve appear in Aretxaga \& Terlevich (1994) and Aretxaga, Cid Fernandes \& 
Terlevich (1997). In this scenario, there are not a supermassive black hole and an 
accretion disc associated with it. From a large sample of AGNs, Hawkins (2002) ruled out 
a pure starburst model. On the other hand, a picture including only a starburst nucleus 
is not suitable for QSO 0957+561, since Kawaguchi et al. (1998) claimed that the 
measured slope of the first-order structure function is in clear disagreement with the 
model. However, QSO 0957+561 could be powered by several independent mechanisms (e.g., 
a nuclear accretion disc together with a circumnuclear stellar region), so SN/SNR events 
must be not discarded as candidates to justify some fluctuations in the light curves of 
the system. In fact, Goicoechea (2002) suggested that two prominent $g$-band 
fluctuations in the QSO 0957+561 light curves may be consistent with circumnuclear 
starburst events. The suggestion was exclusively based on the time delay distribution 
and rough time-energy criteria, and the shape of the events was not taken into account. 
In this section, we are going to properly discuss the starburst origin of the two 
prominent APO $g$-band variations as well as the two GLITP $V$-band gradients. We note
that the existence of a circumnuclear stellar region including young stars, starburst
activity and so on, is also supported by another recent work. Apart from the central
(nuclear) far UV emission, Hutchings (2003) reported the presence of a circumnuclear 
far UV emission within a radius of 0.3 arcsec.

The SN/SNR luminosity is dominated by the standard SNR peak (e.g., Aretxaga, Cid 
Fernandes \& Terlevich 1997; Kawaguchi et al. 1998). The rise from $t = 0.3 t_{sg}$ to 
$t = t_{sg}$ and the subsequent power-law decline define the SNR event, which is the 
main feature of the total emission. The initial SN flash (steep rise at $t = 0$ and 
decay at $t > 0$) and the peaks associated with cooling instabilities are secondary 
features. As usual, $t_{sg}$ is the time when the SNR reaches the maximum of its 
radiative phase. We only consider main events (i.e., SNR ones) and test the possible 
relation between SNR activity and several fluctuations in the records of QSO 0957+561 
(see here above). First of all, from the light curve of one image in an optical band, 
it is deduced the background magnitude. If the central wavelength of the filter's 
bandpass is $\lambda_0$, then we derive $m_{back}(\lambda_0)$. Using standard laws (e.g., 
L\'ena, Lebrun \& Mignard 1998), this background magnitude can be converted to a 
monochromatic flux $F_{back}(\lambda_0)$. We note that the observed background flux at 
$\lambda_0$ is emitted at a shorter wavelength $\lambda = \lambda_0/(1 + z_s)$. Secondly,
$m(\lambda_0) = m_{back}(\lambda_0) - 2.5$ log$[F(\lambda_0)/F_{back}(\lambda_0)]$, where
$F(\lambda_0) = F_{back}(\lambda_0) + F_{SNR}(\lambda_0)$. Hence, in a direct way, we 
obtain the relationship $m(\lambda_0) = m_{back}(\lambda_0) - 2.5$ log$(1 + f)$, with $f =
F_{SNR}(\lambda_0)/F_{back}(\lambda_0)$. Therefore, the fluctuation induced by the SNR 
activity depends on the ratio betweeen the observed SNR flux and the observed background
flux. Thirdly, to account for the observed SNR flux, we must use the cosmological law 
$\lambda_0 F_{SNR}(\lambda_0) = \tau_{SNR} [\lambda L_{SNR}(\lambda)]/(4\pi D_L^2)$,
where $L_{SNR}(\lambda)$ is the monochromatic luminosity, $\tau_{SNR}$ is the 
extinction-magnification factor and $D_L$ is the luminosity distance. During the cooling 
epoch at $t \ge t_{sg}$ (e.g., Shull 1980; Terlevich et al. 1992), a luminosity 
$L_{shock}$ is emitted outward by the forward shock. Most of the energy is radiated in 
the far UV and X-rays, but the initial spectrum could be distorted from the interaction 
with the unperturbed circumstellar material. In general, the emergent luminosity will be
less than $L_{shock}$. An important reprocessing occurs by means of the interaction 
between the radiation emitted inwards (with a luminosity $L_{shock}$) and the outer 
ultradense shell. Half of the reprocessed luminosity (in the outer shell) is reemitted
outward, and it should also cross the unperturbed circumstellar medium. Therefore, 
$\lambda L_{SNR}(\lambda) = \epsilon_{nuv} (2L_{shock})$, where $\epsilon_{nuv}$ is the
near UV efficiency, or equivalently, the ratio between the emergent near UV luminosity
and the total luminosity by the forward shock. We consider the emission of near UV light
($\lambda \approx$ 2100--2300 \AA), because we deal with observations in the $g$ 
($\lambda_0 \approx$ 5067 \AA) and $V$ ($\lambda_0 \approx$ 5500 \AA) bands. All the 
estimations are made in a cosmology with $\Omega_{\Lambda}$ = 0, $\Omega_M$ = 1 and $H$ 
= 66 km s$^{-1}$ Mpc$^{-1}$. 

The key function $f$ has a general form: $C(x/t_{sg0})[(t_0 - 0.3 t_{sg0})/0.7 t_{sg0}]$
at $0.3 t_{sg0} \le t_0 \le t_{sg0}$, and $C(x/t_{sg0})(t_{sg0}/t_0)^{11/7}$ at $t_0 \ge 
t_{sg0}$. The times are not rest-frame ones, but times measured by the observer at 
redshift zero. A constant factor $C$ includes cosmological effects, the magnification of 
the light by the lens and so on, whereas the parameter $x = \epsilon E_{51}$ is related 
to the energy of the SNR in units of 10$^{51}$ erg ($E_{51}$) and a global efficiency
$\epsilon = \epsilon_{dust} \epsilon_{nuv} <$ 1. The global efficiency incorporates the 
extinction by dust in the host and lens galaxies and the Milky Way ($\epsilon_{dust}$). 
We begin with the analysis of the APO $g$-band variations in the light curve of 
Q0957+561B (Kundi\'c et al. 1997). In this first study, the $C$ value is of 1. It is not 
difficult to roughly fit the two prominent observed peaks. In Figure 10 we show both the 
observed trends (filled squares) and the fits (open circles). We do not fit the secondary
features that appear just before each peak (for example, see the behaviour between the 
days 1050 and 1080), which could be associated with the SN explosions. For the main peak 
around day 1105, the SNR parameters are $x$ = 6 and $t_{sg0}$ = 25 days, while for the 
peak around day 1180, the parameters are $x$ = 3 and $t_{sg0}$ = 30 days. The $x$ values 
are in agreement with high energies $E \ge$ 10$^{52}$ erg. These energies slightly exceed
the expected ones for type II SNe and are similar to the energy released in the 
explosions of hypernovae. However, for a well-followed-up type II supernova (SN 1988Z), 
there is evidence in favor of a total radiated energy close to 10$^{52}$ erg (Aretxaga 
et al. 1999). From the time-scales $t_{sg0} \approx$ 25--30 days, we infer a rest-frame 
characteristic time of $t_{sg} \approx$ 10 days. As $t_{sg}$ (days) = 230 $E_{51}^{1/8}
n_7^{-3/4}$, where $n_7$ is the circumstellar density in units of 10$^7$ cm$^{-3}$, one
can easily find a density of $n \ge$ 10$^9$ cm$^{-3}$. This high density is not so
surprising (e.g., Filippenko 1989), and the energy and environmental density values do
not permit to rule out the hypothesis of SNR activity. 

\begin{figure}
\psfig{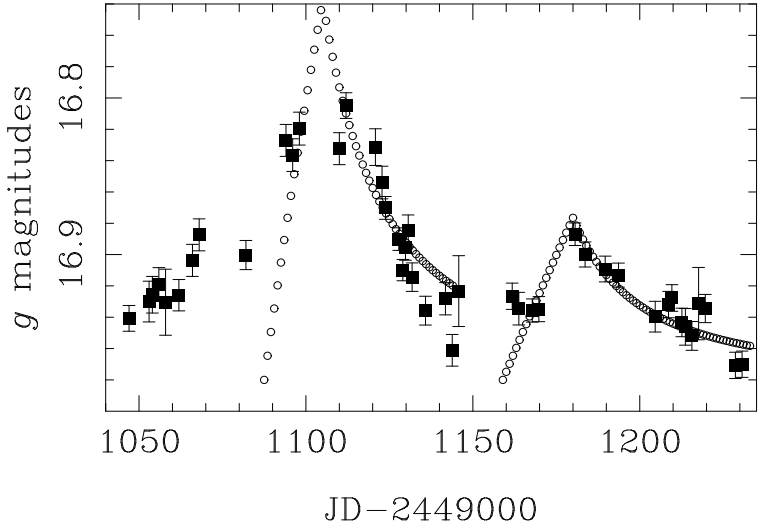}  
\caption{Light curve of Q0957+561B obtained at the Apache Point Observatory during the 
1996 season. The filled squares represent the photometric data in the $g$ band, while 
the open circles correspond to SNR fits. If the two prominent peaks are caused in a 
starburst scenario, high energies ($E \ge$ 10$^{52}$ erg) and environmental densities 
($n \ge$ 10$^9$ cm$^{-3}$) are involved.}
\label{Fig. 10}
\end{figure}

\begin{figure}
\psfig{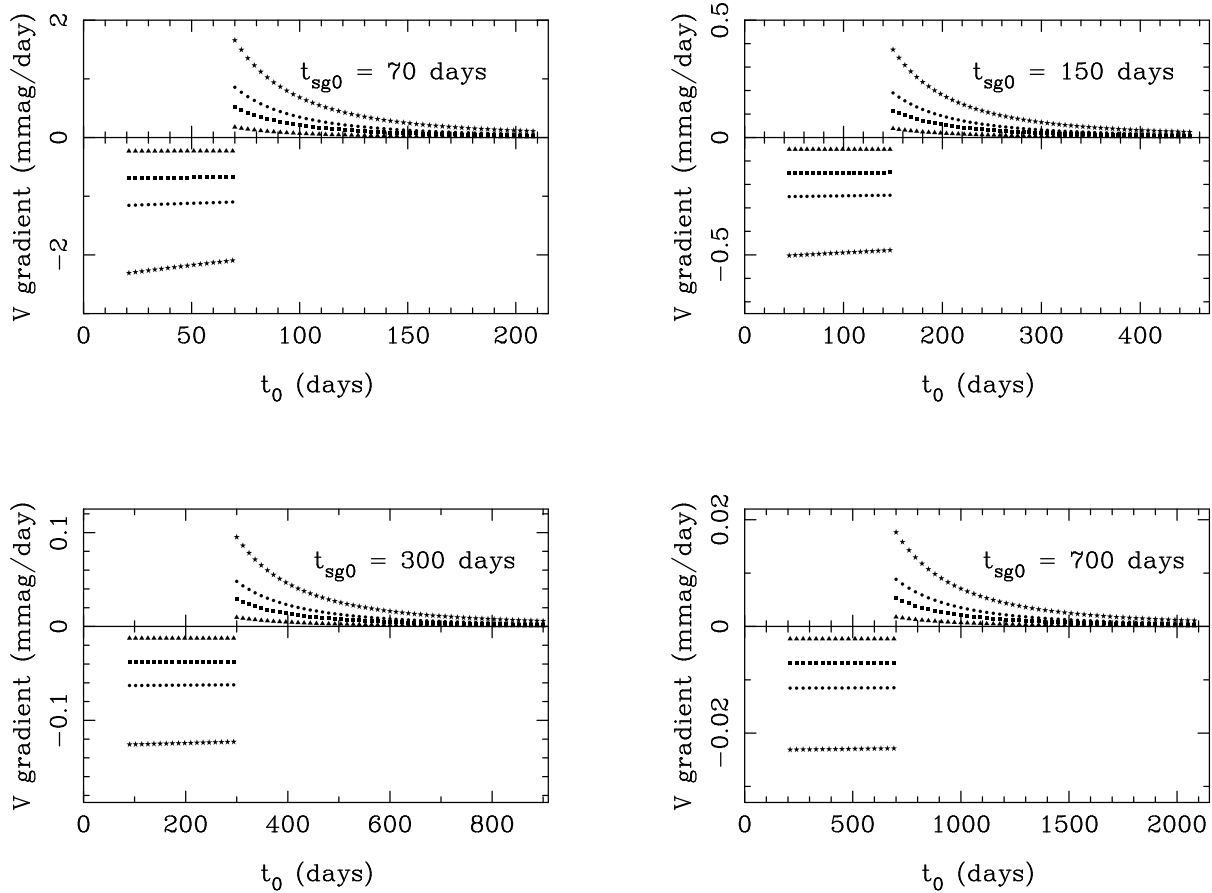}
\caption{$V$-band gradients due to SNR activity. We simulate rises for Q0957+561A 
(left-hand of the panels) and declines for Q0957+561B (right-hand of the panels). For
each characteristic time, we take four values of $x$: 0.25 (triangles), 0.75 (squares), 
1.25 (circles) and 2.5 (stars).}
\label{Fig. 11}
\end{figure}

On the other hand, in the previous section, we report that the GLITP $V$-band photometry 
has two well-defined gradients: one rise in Q0957+561A and one decline in Q0957+561B. 
Are these gradients roughly consistent with SNR activity?. We try to answer this query 
from SNR simulations. In the new analysis, we focus on the time derivatives (gradients) 
$dm(\lambda_0)/dt_0$ (mmag/day) = $-$ 1086 $(1 + f)^{-1}(df/dt_0)$. The GLITP gradients 
have a duration of about 45 days, so we only consider $t_{sg0}$ values longer than 50 
days. In Figure 11 we present the results of the simulations: rises for the A component
($C$ = 2.92) and declines for the B component ($C$ = 2.04). We take four representative 
values of $x$, i.e., $x$ = 0.25 (triangles), $x$ = 0.75 (squares), $x$ = 1.25 (circles) 
and $x$ = 2.5 (stars). With respect to the declines (Q0957+561B), we cannot fairly 
reproduce a constant fall of 0.3--0.4 mmag/day during a time interval of 45 days.
For $t_{sg0} >$ 150 days, the declines are significantly smaller than the observed 
one, while for shorter $t_{sg0}$ values, a 45 days decline of about 0.35 mmag/day do 
not seem to be plausible. With respect to the rises (Q0957+561A), the observed gradient 
is consistent with a SNR scenario. A 45 days rise of about 0.9 mmag/day may be explained
from the simulations in the left-hand top panel of Fig. 11. However, we can use some 
complementary information on the feature in Q0957+561A. Oscoz et al. (2002) reported a
large brightness enhancement of the Q0957+561A component during the 2000 and 2001
monitoring campaigns with the IAC-80 telescope (see Figs. 1 and 2 in that paper). They
found a 250 mmag variation lasting 500 days, so that the 45 days GLITP rise corresponds
to the beginning of the brightness enhancement. Assuming a 500 days rise of about 
0.5 mmag/day (an average slope), the brightness increase must be due to a physical 
phenomenon different to SNR activity (see the right-hand bottom panel of Fig. 11). Indeed 
the GLITP+IAC data suggest the existence of two non-starburst gradients in the light 
curves of QSO 0957+561. 

\section{Conclusions}

We have observed the lensed quasar QSO 0957+561 during two months in 2000, using the
StanCam/NOT instrument. From the daily $VR$ images and PSF photometry, we inferred 
very detailed $VR$ light curves of the two components A and B. The final brightness 
records are characterized by the following properties:
\begin{enumerate}
\item In each optical band, the daily variability is similar in the two QSO components. 
We justified that the correlated day-to-day fluctuations are caused by observational 
(systematic + random) noise. The peaks of noise could be mainly due to the mismatch 
between the reference PSF and the PSF in the frame region where is placed the lens 
system. While the typical $R$-band uncertainties are of about 5 mmag (Q0957+561A) and 6 
mmag (Q0957+561B), the typical $V$-band errors are of $\approx$ 9$-$10 mmag.
\item In each optical band, there are uncorrelated linear gradients on a six weeks
time-scale. For the A component, we measured rises of $-$ 0.72 $\pm$ 0.08 mmag/day (in
the $R$ band) and $-$ 0.89 $\pm$ 0.12 mmag/day (in the $V$ band), whereas for the B
component, we derived decreases of + 0.20 $\pm$ 0.09 mmag/day ($R$-band) and + 0.36 
$\pm$ 0.13 mmag/day ($V$-band). There is some evidence for chromatic variability, but 
the $V$-band gradients are marginally consistent with the $R$-band ones (using 1$\sigma$
confidence intervals).
\end{enumerate}

As quoted in the previous paragraph, the observed $V$-band gradients have values below
one millimagnitude per day and a duration of about 45 days. We discussed a possible 
physical scenario to account for these features: supernova remnant (SNR) activity. Two
prominent $g$-band variations reported by Kundi\'c et al. (1997) are roughly consistent
with the existence of a circumnuclear starburst region, so the starburst origin of the
new $V$-band gradients is an attractive possibility. However, the observed decline (in
Q0957+561B) is in disagreement with SNR simulations. On the other hand, even though an
optimistic result was obtained from the comparison between the observed rise (in
Q0957+561A) and SNR simulations, extended observations of the rise (Oscoz et al. 2002)
led to reject a starburst mechanism. Therefore, we finally conclude that the new 
$V$-band gradients must be associated with physical phenomena different to SNR activity.

\section*{Acknowledgments}

We would like to thank Jan-Erik Ovaldsen for interesting comments on variability and 
time delays in QSO 0957+561. We also thank the anonymous referee for useful comments.
The GLITP observations were made with the Nordic Optical Telescope (NOT), which is 
operated on the island of La Palma jointly by Denmark, Finland, Iceland, Norway, and 
Sweden, in the Spanish Observatorio del Roque de Los Muchachos of the Instituto de 
Astrof\'{\i}sica de Canarias (IAC). We are grateful to the technical team of the 
telescope. AU thanks the Spanish Department of Science and Technology (MCyT) for two 
fellowships. JAM is a {\it Ram\'on y Cajal Fellow} from the {\it Ministerio de Ciencia 
y Tecnolog\'{\i}a} of Spain. This work was supported by the P6/88 project of the IAC, 
Universidad de Cantabria funds, and the MCyT grants AYA2000-2111-E and AYA2001-1647-C02.

\bsp

\end{document}